\documentclass[conference]{IEEEtran}
\IEEEoverridecommandlockouts
\usepackage{cite}
\usepackage{amsmath,amssymb,amsfonts}
\usepackage{algorithmic}
\usepackage{graphicx}
\usepackage{textcomp}
\usepackage{xcolor}

\usepackage[caption=false,font=footnotesize]{subfig}

\def\BibTeX{{\rm B\kern-.05em{\sc i\kern-.025em b}\kern-.08em
    T\kern-.1667em\lower.7ex\hbox{E}\kern-.125emX}}
\begin{document}

\title{A Case for Quantifying Statistical Robustness of Specialized Probabilistic AI Accelerators
\thanks{This work appears as a poster in 2019 IBM IEEE CAS/EDS - AI Compute Symposium, Yorktown Heights, NY, Oct. 2019.}
}

\author{
\IEEEauthorblockN{Xiangyu Zhang, Sayan Mukherjee, Alvin R. Lebeck} \\
\IEEEauthorblockA{\textit{Duke University}
}
}

\maketitle
\section{Introduction}

Statistical machine learning, like other methods in artificial intelligence, has become an important type of workload for computing systems. Such workloads often utilize probabilistic algorithms, which enable the potential to provide generalized frameworks to solve a wide range of problems, such as computer vision, healthcare, robotics, and computational biology. 
As alternatives to Deep Neural Networks, these algorithms provide easier access to interpreting why a given result is obtained. 
The critical section of probabilistic algorithms, such as Markov Chain Monte Carlo (MCMC) method, is iteratively generating samples from parameterized distributions, which takes hundreds of CPU cycles even for simple distributions.

To address this inefficiency, accelerators are proposed using specialized hardware. These accelerators often improve the hardware efficiency by using some approximation techniques, such as reducing bit representation, truncating small values to zero, or simplifying the Random Number Generator (RNG). 
Understanding the influence of these approximations on result quality is crucial to meeting the quality requirements of real applications. 
A common approach is to compare the end-point result quality using community-standard benchmarks and metrics, such as accuracy in classification, end-point error in optical flow, and bad-pixel percentage (BP) in stereo vision. 
Although important, merely evaluating the end-point result quality on a probabilistic architecture is insufficient since ground truth data is not always available. 
Therefore, \textit{a probabilistic architecture should provide some measure (or guarantee) of statistical robustness.}

This work takes a step towards quantifying the statistical robustness of specialized hardware MCMC accelerators. We start by asking the following questions compared to software-only results: 
1) \textit{What sampling quality can the accelerator provide?}, 2) \textit{How many more iterations does the accelerator need to converge?}, 3) \textit{How different are the results of the accelerator?} 
To answer these questions, we propose an empirical method with three pillars of statistical robustness: \textbf{sampling quality}, \textbf{convergence diagnostic}, and \textbf{goodness of fit}. 
Each pillar has at least one quantitative metric. The method utilizes application data without the need to know the ground truth. 
We apply this method to analyze the statistical robustness of an MCMC accelerator \cite{zhang2018isca}, with some modifications, using stereo vision as a case study. The method also applies to other probabilistic accelerators. 
We also use an empirical method for data-independent analysis to measure the distribution divergence between the accelerator and the 64-bit floating-point software. This method provides the worst-case divergence of the MCMC accelerator and helps reveal potential design problems. Both methods can be used in design space exploration. 

\begin{figure*}[t]
\centering
	\captionsetup[subfigure]{width=0.30\textwidth}
	\subfloat[Mean ESS overall and in the active region]{
		\includegraphics[width=0.30\textwidth,keepaspectratio,trim=0mm 0mm 0mm 0mm]{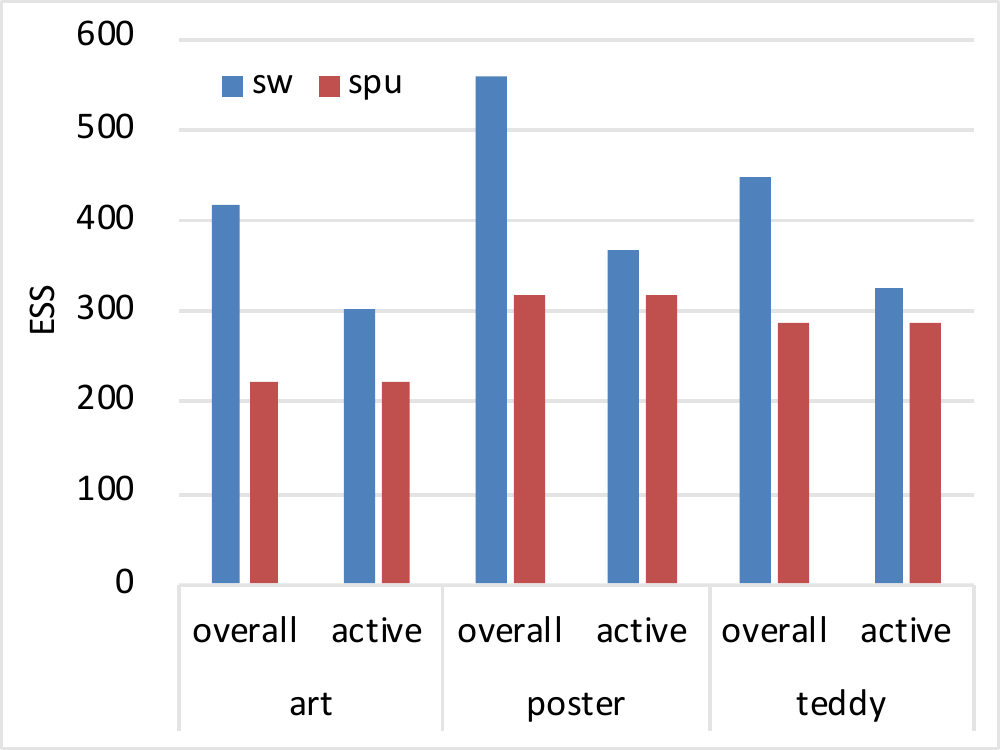}
		\label{fig:ess}
    }
	\subfloat[Convergence percentage based on $\hat{R}$]{
		\includegraphics[width=0.29\textwidth,keepaspectratio,trim=0mm 0mm 0mm 0mm]{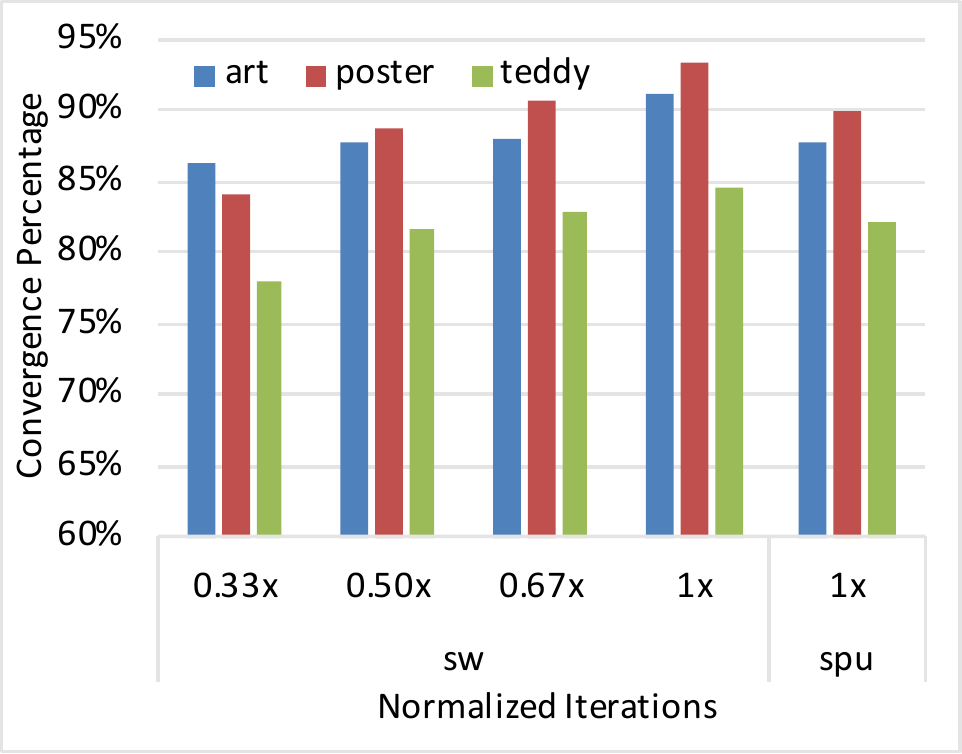}
		\label{fig:rhat}
    }
	\subfloat[Box plot of $R^2$ comparison]{
		\includegraphics[width=0.30\textwidth,keepaspectratio,trim=10mm 10mm 15mm 10mm]{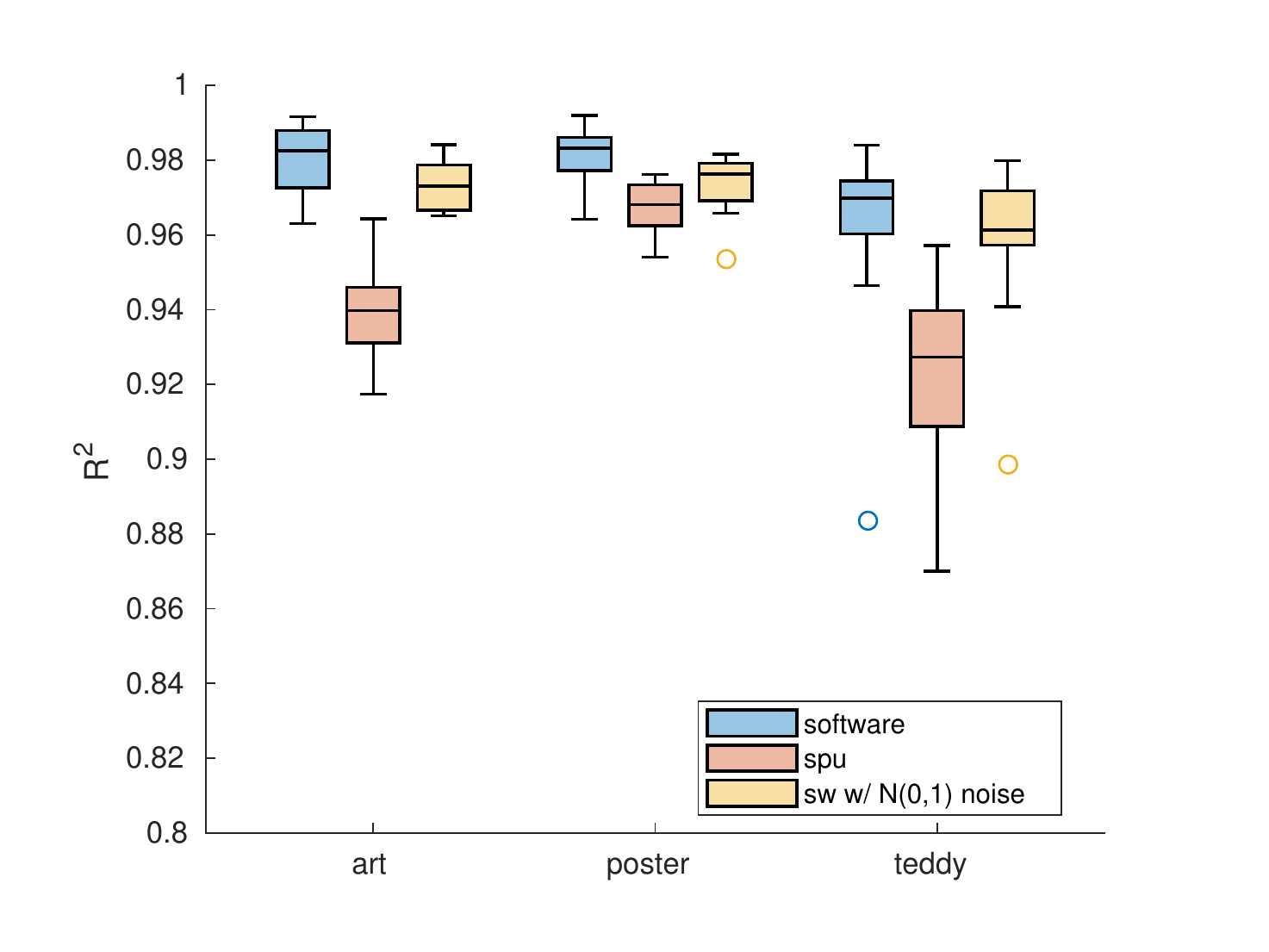}
		\label{fig:r2}
    }
	\caption{Statistical robustness results on SPU under pure-sampling method. }
\label{fig:three_pillar}
\end{figure*}
\section{Methodology}

\subsection{Three Pillars of Robustness}
MCMC is a theoretically important and powerful 
method for solving the inference problem by iteratively sampling random variables (RVs) and ultimately converging to the target distribution. In addition to end-point result quality, we claim the following statistical properties of the accelerator need to be evaluated compared to the 64-bit FP software runs. 

\paragraph*{Pillar 1) Sampling Quality} 
A sampling algorithm with perfect sampling quality generates independent samples. However, the dependent nature of MCMC creates autocorrelation between samples and effectively reduces the number of independent samples being drawn. Although sample quality is influenced by the algorithm, it can be further reduced by hardware approximations. 
The notion of Effective Sample Size (ESS)
is commonly-used to measure how many effective independent samples is drawn in the dependent samples from a RV. Multiple methods are available to estimate ESS. We estimate ESS on a univariate RV by autocorrelation function \cite{liu2015}. In our case study, we report the mean ESS of RVs (pixels in the image) as a singular metric.

\paragraph*{Pillar 2) Convergence Diagnostic}
An important question for MCMC is when it converges. Similar to sampling quality, hardware approximation can influence convergence. This can be measured using Gelman-Rubin's $\hat{R}$, a convergence diagnostic that indicates whether the MCMC runs on a univariate RV converge at a certain iteration by comparing the between-chain variance ($B$) and with-in chain variance ($W$) across multiple independent runs. 
The formula of $\hat{R}$ is given by Brooks et al. \cite{brooksgelman98}. As a rule of thumb, a univariate RV is considered converged when $\hat{R}<1.1$. 
In our case study, a univariate RV (a pixel) can have $W=0$ both in software and hardware runs, where the original metric has no definition. We define the univariate
RV is converged when $B=0$ and $W=0$, and not converged when $B>0$ and $W=0$. We propose convergence percentage, the percentage of univariate RVs that have $\hat{R}<1.1$ or $B,W=0$, as a new singular metric. 

\paragraph*{Pillar 3) Goodness of Fit} Ideally, we need a metric for how close results are between software and hardware. However, due to the probabilistic nature of MCMC, each MCMC run can have different end-point results, either in software or hardware. To account for this variation, we use the ``goodness of fit'' metric $R^2$ to measure how well a MCMC run result fits to a reference obtained from the mode of all software runs. The statistics (e.g. mean and variance) of $R^2$s across multiple hardware and software runs indicate how close the results are in hardware to the software. We propose that a Kolmogorov-Smirnov (KS) permutation test can be conducted on $R^2$s, if it is necessary to determine whether $R^2$s in software and hardware are drawn from the same distribution.

\subsection{Distribution Divergence}
Approximation techniques in hardware, such as limited precision and truncation, introduces divergence from the distribution obtained from 64-bit FP software. 
Quantifying the distribution divergence between software $D_{sw}$ and hardware $D_{hw}$ provides insights on why the results are good (or bad) and gives the worst-case divergence. 
Given that KL-divergence goes to infinity when $D_{hw}$ contains  
zeros, we use Jensen-Shannon Divergence as the divergence measurement. 

\section{A Case Study}
We apply the above methods to a Stochastic Processing Unit (SPU), an MCMC accelerator based on the accelerator proposed by Zhang, et al. \cite{zhang2018isca} with modifications to replace the non-CMOS sampler with a CMOS discrete sampler using a 19-bit LFSR. The SPU supports both a pure-sampling method and sampling with simulated annealing, an optimization algorithm that converges faster to a point solution. 
We report the results of the pure-sampling method in this work. We select three input data for the stereo vision application. Using 20 runs of each input with the same number of iterations, the means for end-point result quality in bad-pixel percentage (BP, lower is better) compared to the ground truth are: \textit{art} 27.6\% in software vs 32.0\% in SPU, \textit{poster} 11.0\% vs 11.5\%, and \textit{teddy} 27.1\% vs 28.5\%. 
Although the BP results differ, no obvious differences are observed visually between software and the SPU in the disparity maps. 
Results for the three pillars of statistical robustness are shown in Fig \ref{fig:three_pillar}. Due to approximations, more RVs in the SPU have zero variance than in software. These RVs have consistently small variance in software, so we consider them as the inactive region and the remaining RVs as active. ESS, shown in Fig \ref{fig:ess}, is collected from the last 1,000 iterations in a single MCMC run per input data. 
We verify that different runs have small ESS difference ($<6$). 
Mean ``overall'' ESS eliminates RVs with zero variance in software and the SPU, respectively. This introduces the bias towards the software since RVs with small but non-zero variances tend to have meaninglessly high ESS.
Thus, we report the mean ESS in the active region where ESS is meaningful. The SPU with 1.1-1.4$\times$ iterations reach the same ``active'' ESS as software. Fig \ref{fig:rhat} shows the convergence percentage comparison. 
The numbers of iterations are normalized with respect to SPU runs. The SPU reaches the same or better convergence percentage than software with 2$\times$ iterations. 
This indicates the SPU needs to be at least 2$\times$ faster (previous work showed the speedups of at least 2.8-5.5$\times$ and up to 84$\times$). 
Fig \ref{fig:r2} compares the $R^2$s for the SPU to that for software and software with input noise $\mathcal{N}(0,1)$ for 20 runs, each with the same number of iterations. The SPU has a slightly lower $R^2$s than software, indicating the differences in BP results. The input noise helps quantify the result differences between software and the SPU by modeling input noise in software. The $R^2$s are consistent with BP result quality. We can use $R^2$ to compare the end-point result quality when ground truth data are not available. 

\section*{Acknowledgements}
Special thanks to Ramin Bashizade and to Intel for supporting this work.
\bibliographystyle{IEEEtran}
\bibliography{ref}

\end{document}